\documentclass[a4paper]{article}
    
    \usepackage[english]{babel}
    \usepackage[utf8x]{inputenc}
    \usepackage[T1]{fontenc}
    \usepackage{cite}

    \usepackage[a4paper,top=3cm,bottom=2cm,left=3cm,right=3cm,marginparwidth=1.75cm]{geometry}
    
	\usepackage{amsmath,amsfonts,amssymb}
	\usepackage{graphicx}
    \usepackage[colorinlistoftodos]{todonotes}
	\usepackage[colorlinks=true, allcolors=blue]{hyperref}

    \usepackage{titlesec,upgreek,adjustbox}
    \newcommand{\mum}{\,\mathrm{\upmu m}} 
    
    \usepackage{tabularx,multirow}
    \usepackage{ulem}
    
    \usepackage{rotating}
    \usepackage{hhline}
    
    \usepackage{ccaption}
    
    \usepackage{array}
	\newcolumntype{L}[1]{>{\raggedright\let\newline\\\arraybackslash\hspace{0pt}}m{#1}}
	\newcolumntype{C}[1]{>{\centering\let\newline\\\arraybackslash\hspace{0pt}}m{#1}}
	\newcolumntype{R}[1]{>{\raggedleft\let\newline\\\arraybackslash\hspace{0pt}}m{#1}}
    
    \usepackage{mathtools}
    \DeclarePairedDelimiter\abs{\lvert}{\rvert}%

	\newcommand{\NA}{\mathrm{NA}}

	\title{Light-sheet microscopy with attenuation-compensated propagation-invariant beams}

	\author{Jonathan Nylk$^{1,}$\footnote{email: jn78@st-andrews.ac.uk} , Kaley McCluskey$^{1,}$\footnote{Present address: Department of Bionanoscience, Kavli Institute of Nanoscience, Delft University of Technology, Van der Maasweg 9 2629HZ Delft, The Netherlands
} , Miguel A. Preciado$^{1,}$\footnote{Present address: School of Physics and Astronomy, University of Glasgow, Glasgow, G12 8QQ, UK} ,\\
    Michael Mazilu$^{1}$, Frank J. Gunn-Moore$^{2}$, Sanya Aggarwal$^{3}$,\\
    Javier A. Tello$^{3}$, David E. K. Ferrier$^{4}$, and Kishan Dholakia$^{1}$\\
    $^{1}$SUPA, School of Physics and Astronomy, University of St Andrews,\\North Haugh, St Andrews, KY16 9SS\\
    $^{2}$School of Biology, University of St Andrews, North Haugh,\\ St Andrews, KY16 9SS\\
    $^{3}$School of Medicine, University of St Andrews, North Haugh,\\ St Andrews, KY16 9SS\\
    $^{4}$Scottish Oceans Institute, Gatty Marine Laboratory,\\School of Biology, University of St Andrews, East Sands,\\St Andrews, KY16 8LB}
 	\date{}
	\begin{document} 
	\maketitle

	\begin{abstract}
	 
	Scattering and absorption limit the penetration of optical fields into tissue, but wavefront correction, often used to compensate for these effects, is incompatible with wide field-of-view imaging and complex to implement. We demonstrate a new approach for increased penetration in light-sheet imaging, namely attenuation-compensation of the light field. This tailors an exponential intensity increase along the illuminating propagation-invariant field, enabling the redistribution of intensity strategically within a sample. This powerful yet straightforward concept, combined with the self-healing of the propagation-invariant field, improves the signal-to-background ratio of Airy light-sheet microscopy up to five-fold and the contrast-to-noise ratio up to eight-fold in thick biological specimens across the field-of-view without any aberration-correction. This improvement is not limited to Airy beam light-sheet microscopy, but can also significantly increase the imaging capabilities of Bessel and lattice light-sheet microscopy techniques, paving the way for widespread uptake by the biomedical community.
    
	\end{abstract}

    \begingroup
	\sloppy

	\section{Introduction}
	\label{sec:intro}
    
    The optical analysis of living tissues and organisms has yielded fascinating insights into biological processes. In particular, light-sheet microscopy (LSM) is revolutionising such studies due to the highly-parallel, high-contrast, wide-field, and minimally phototoxic nature of this imaging method \cite{Huisken2004,Ahrens2013,Yang2016,Hockendorf2012}. However, the penetration of light into such specimens is ultimately limited by the heterogeneous nature of tissue, making observation within deep tissues problematic. A suite of adaptive optical methods have been developed for optical imaging. They typically address the problem by selecting a `guide star' and exploiting the orthogonality of input probe fields to correct for aberrations in a localised area around the guide star\cite{Ji2017}. While such aberration correction has yielded impressive results in microscopy \cite{Ji2017}, the limited region of improvement around the corrected point is discordant with the wide field-of-view (FOV) encountered in most LSM systems \cite{Power2017}. Indeed, aberration-correction has been largely unexplored in LSM aside from a few limited attempts \cite{Dalgarno2012,Bourgenot2012,Masson2015,Wilding2016}. A step change for this field would be to implement an all-optical approach to overcome the detrimental effects of tissue scattering and absorption on the incident light field across a wide FOV, without the need for adaptive optics. This is what we present here with the use of attenuation-compensated propagation-invariant fields.
    
    Propagation-invariant light fields, most notably Airy \cite{Vettenburg2014,Yang2014,Piksarv2017} and Bessel \cite{Fahrbach2010,Planchon2011,Olarte2012,Chen2014,Gao2012} beams, have gained interest in LSM primarily for their quasi-non-diffracting properties, maintaining a narrow transverse profile over distances much larger than the Rayleigh range of an equivalent Gaussian beam and enabling high-resolution imaging over an increased FOV. The `self-healing' ability of propagation-invariant beams has also been particularly advantageous for imaging. Such self-healing beams are able to recompose their transverse profile rapidly on propagation after being partially blocked by an obstacle\cite{Bouchal1998,Broky2008,Mazilu2010,Chu2012,Zhang2015}, which has translated to a resistance against aberration in turbid media, with the beam profile degrading at a reduced rate compared to a Gaussian beam \cite{Fahrbach2012,Chen2015,Nylk2016}. While these fields maintain their profile in the presence of scattering, intensity losses associated with both scattering and absorption still pose a challenge for deep tissue imaging.
    
    Importantly, the longitudinal intensity envelope of a propagation-invariant field is determined by the method of its generation \cite{Cizmar2009}, and recent studies have shown that the envelope can be tailored arbitrarily \cite{Cizmar2009,ZamboniRached2004,Preciado2012,Preciado2014,Schley2014}. Here we exploit such control of the longitudinal envelope to counteract attenuation of the illuminating light-sheet in optically thick specimens, in order to recover high-contrast images from deep tissue layers without increased irradiation of superficial tissues. This attenuation-compensation is achieved by tailoring an exponential increase of intensity along the beam propagation with an exponent matched to the decrease in intensity caused by attenuation.
    
    We focus upon attenuation-compensated Airy LSM to demonstrate our principles, and show deep tissue imaging of \textit{Spirobranchus lamarcki} opercula and sections of mouse brain. In all experiments, we observed a marked increase in signal-to-background ratio (SBR) and contrast-to-noise ratio (CNR), with many indiscernible features being raised noticeably above the noise floor, in turn leading to an extension of the usable FOV by up to $100\%$ (corresponding to an increased depth penetration of $150\mum$).

    \section{Results}
	\label{sec:results}
    
    \subsection{Theoretical study of attenuation/attenuation-compensation on image performance}
	\label{subsec:TheorySimsResults}
    
    	The intensity of a beam propagating through a linearly attenuating medium (for simplicity we consider pure absorption; $C_{attn} \equiv C_{abs}$) will decrease exponentially as given by the Beer-Lambert law (Fig. \ref{fig:overview}(a,b)). For propagation-invariant beams, there is a mapping between the transverse coordinate in the pupil plane and the longitudinal coordinate near the focal plane\cite{Cizmar2009,ZamboniRached2004,Preciado2012,Preciado2014,Schley2014}. Therefore, by judicious choice of weighted complex amplitudes for different wave vectors of the beam, the intensity profile can be engineered such that the net attenuation is zero (Fig. \ref{fig:overview}(c)). The cylindrical pupil function for an attenuation-compensated Airy light-sheet is given by:
    
    \begin{equation}
    \label{eq:MainTextAiryPupil1D}
    	P(u) = A_{\sigma} \exp(2\pi i \alpha u^{3}) \exp(-u^{8}) \exp(\sigma[u-1])H(\sqrt{2}-\abs{u})
    \end{equation}
    
    \noindent where $u$ is the normalised pupil coordinate corresponding to the $z-$axis of the microscope, $A_{\sigma}$ is a real scaling factor, $\alpha$ dictates the propagation-invariance of the Airy light-sheet \cite{Vettenburg2014}, $\sigma$ dictates the degree of linear attenuation-compensation, $H(\cdot)$ denotes the Heaviside step function, and the light-sheet propagates in the positive $x-$direction (see Supplementary Note S1). Similar mappings exist for Bessel beams (see Supplementary Note S2).
    
    When attenuation-compensation is applied to an Airy light-sheet, the effective attenuation coefficient, $C_{attn}'$, then becomes:
    
    \begin{equation}
    \label{eq:MainTextModifieldAttenuationCoeff}
    	\begin{aligned}
    		C_{attn}' &= C_{attn} - \chi \\
            &= C_{attn} - \frac{10 \sigma \NA^{2}}{3 \pi^{2} n \alpha \lambda}
        \end{aligned}
    \end{equation}
    
    \noindent where $\NA$ is the numerical aperture corresponding to $u=1$, $n$ is the refractive index of the sample, and $\lambda$ is the vacuum wavelength of the illumination (see Supplementary Note S1)\cite{Preciado2014}. For $\chi = C_{attn}$, the main caustic of the Airy light-sheet propagates as if without any loss.
    
    \begin{figure}[t]
      \centering
      \includegraphics[width=\linewidth]{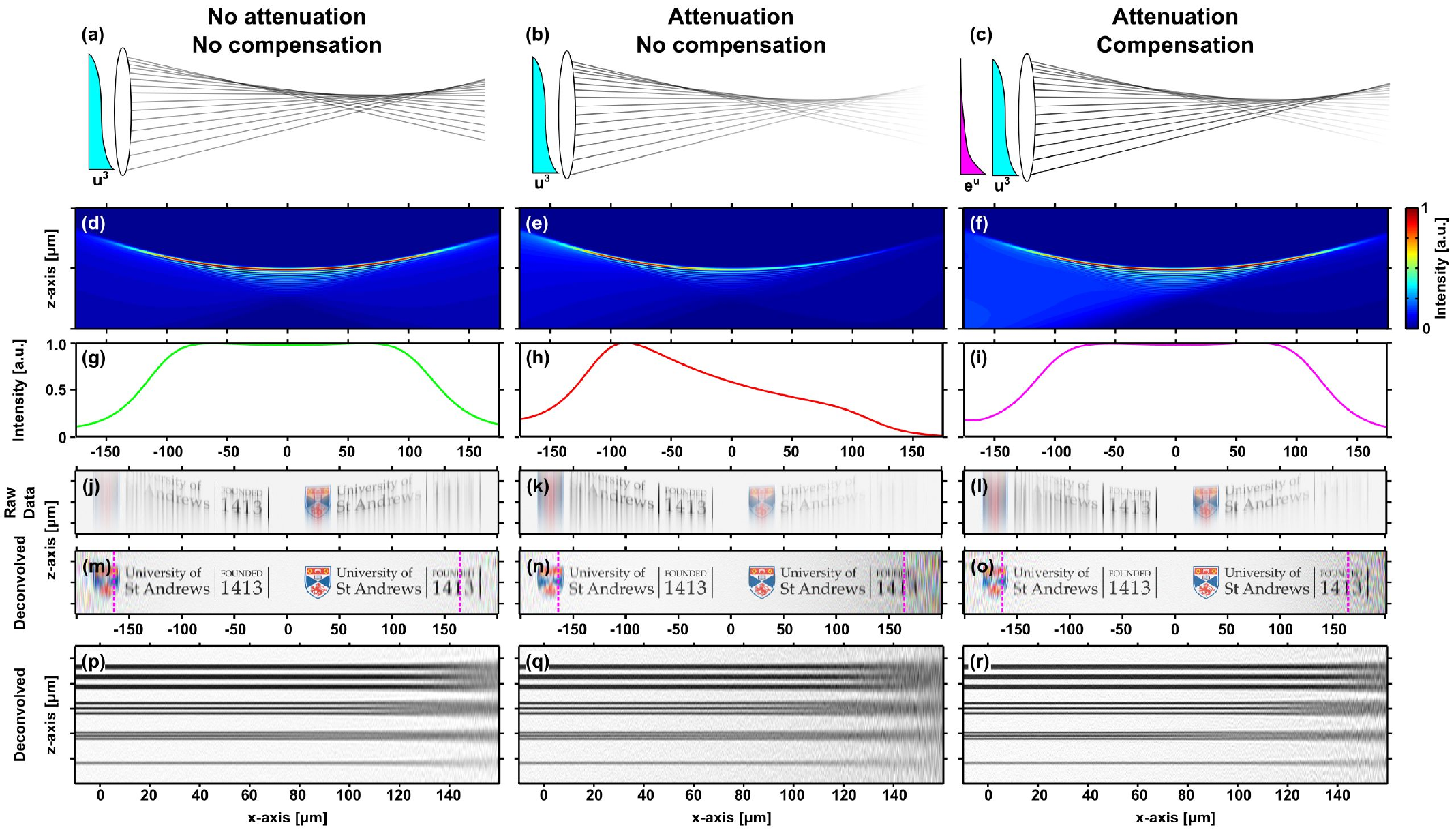}
      \caption{Ray optics representations of Airy light-sheet formation without (a) and with (b) attenuation ($C_{attn} = 64.95$cm$^{-1}$), and with attenuation-compensation (c; $\sigma = 0.54$). (d-f) Wave optical simulations of light-sheet profiles, (g-i) peak transverse intensity as a function of longitudinal coordinate, (j-o) simulated recorded and deconvolved images of the University crest, and (p-r) simulated deconvolved images of a 1D resolution target respectively for the light-sheets shown in (a-c). Pink dashed lines in (m-o) indicate the edge of the FOV from theory. 1D resolution target (p-r) has line width/spacing: $2\mum$ (top), $1\mum$, $0.6\mum$, $0.2\mum$ (bottom). Simulation parameters were set to mirror experimental parameters (see Methods (Section \ref{sec:methods})). These datasets can be accessed at\cite{Data}.}
      \label{fig:overview}
	\end{figure}
    
    The selective delivery of additional power to greater depths within an attenuating medium allows the Airy light-sheet to maintain consistent intensity along its main caustic (Fig. \ref{fig:overview}(d-i)). However, with our experimental parameters (see Methods (Section \ref{sec:methods})), for $\sigma > 0.54$ we found the peak intensity was no longer located on the main caustic and gave unpredictable longitudinal intensity profiles (see Supplementary Note S1). Therefore we defined $\sigma = 0.54$ as the maximum permitted attenuation-compensation for this study, allowing complete compensation up to $C_{attn} = 64.95$cm$^{-1}$ over the whole FOV of the light-sheet as determined from theory ($328\mum$)\cite{Vettenburg2014}.
     
     Airy LSM utilises deconvolution of the recorded images, based on the point spread function (PSF) of the illuminating light-sheet to achieve high axial resolution. Modification of the deconvolution procedure to incorporate the attenuated light-sheet profile (see Supplementary Note S3) facilitates correct re-normalisation of the sample fluorophore distribution, but at the cost of amplified noise in regions where the illumination is weak. Due to the increased intensity at depth when using attenuation-compensation, the signal is restored without noise amplification (Fig. \ref{fig:overview}(j-r), Supplementary Fig. S18 and S19).
     
     As deconvolution is required for Airy LSM, the axial image quality can be explored through analysis of the modulation transfer function (MTF) of the light-sheet profiles. We found that attenuation results in a marked reduction in the support of high spatial frequencies that is restored with the use of attenuation-compensation (see Supplementary Note S4). Interestingly, for complete attenuation-compensation ($C_{attn}'=0$) the bandwidth was found to be higher than for the ideal (non-attenuating) case. Deconvolution requires \textit{a priori} knowledge of the sample attenuation and incorrect estimation of $C_{attn}$ will result in image artefacts. Estimation of $C_{attn}$ to within $10\%$ of the true value was sufficient to minimise the effect of such artefacts (see Supplementary Note S5).
     
     Similarly, for Bessel LSM methods, the use of attenuation-compensation does not affect the transverse beam shape (see Supplementary Note S6) and is therefore compatible will all existing Bessel LSM techniques (e.g. \cite{Fahrbach2010,Fahrbach2012,Zhang2014,Zhang2016,Planchon2011,Chen2014,Gao2012,Olarte2012}).
    
    \subsection{Attenuation-compensated Airy light-sheet microscopy in an attenuating phantom}
	\label{subsec:BeadsResults}
    
   		To experimentally test the performance of attenuation-compensated Airy LSM under strongly attenuating conditions, we developed an attenuation-compensated Airy LSM (see Methods (Section \ref{sec:methods})) and first imaged sub-diffraction-limited fluorescent beads in an absorbing dye solution ($C_{attn}=(55\pm1)$cm$^{-1}$; see Methods (Section \ref{sec:methods}), Supplementary Notes S7 and S8).
        
        Figure \ref{fig:beadExperiment} shows $x-z$ maximum intensity projections of the recorded data (Fig. \ref{fig:beadExperiment}(a,d,g)) and deconvolved images (Fig. \ref{fig:beadExperiment}(b,e,h)). The sample was imaged with three different levels of compensation; no compensation ($\sigma=0$; $C_{attn}'=(55\pm1)$cm$^{-1}$; Fig. \ref{fig:beadExperiment}(a-c)), partial compensation ($\sigma=0.23$; $C_{attn}'=(27\pm1)$cm$^{-1}$; Fig. \ref{fig:beadExperiment}(d-f)), and full compensation ($\sigma=0.46$; $C_{attn}'=(0\pm1)$cm$^{-1}$; Fig. \ref{fig:beadExperiment}(g-i)). Figure \ref{fig:beadExperiment}(c,f,i) show magnified views of the region indicated by the dashed box in Fig. \ref{fig:beadExperiment}(b,e,h).
        
        \begin{figure}[tb]
    	\centering
      	\includegraphics[width=1\linewidth]{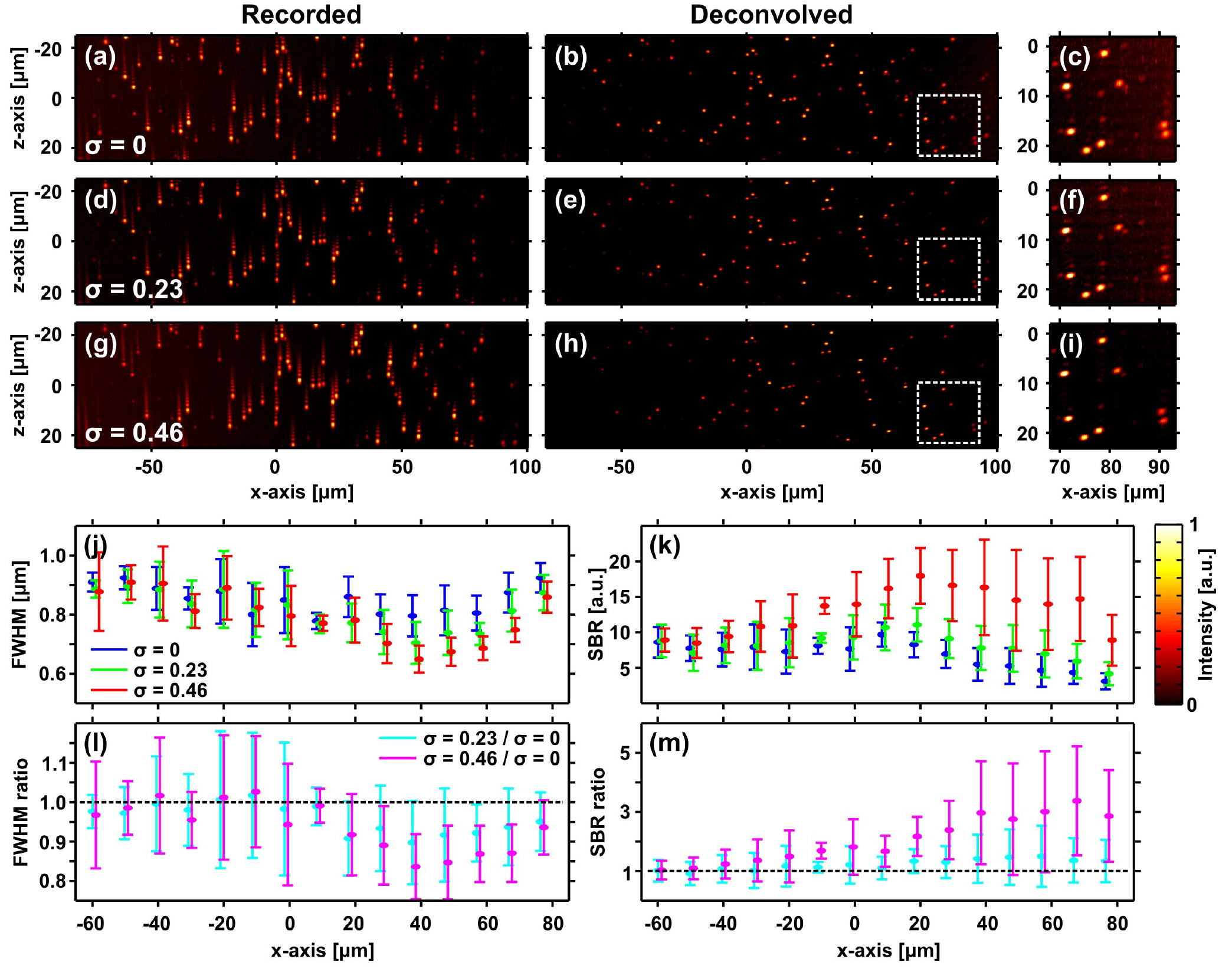}
      	\caption{Maximum intensity projections of recorded data (a,d,g) and deconvolved images (b,e,h) of sub-diffraction limited fluorescent microspheres in an absorbing phantom with $C_{attn} = 55 \pm 1$cm$^{-1}$. (a,b) No attenuation-compensation ($\sigma = 0$), (d,e) $\sigma = 0.23$, (g,h) $\sigma = 0.46$ (full attenuation-compensation). (c,f,i) show zoomed-in views of the region indicated by the dashed box in (b,e,h). (j,k): Axial resolution determined by FWHM of fluorescent microspheres and local SBR respectively as a function of light-sheet propagation (mean $\pm$ st. dev. - $10\mum$ binning); $\sigma = 0$ (blue), $\sigma = 0.23$ (green), $\sigma = 0.46$ (red). (l,m): ratios of the graphs shown in (j,k); $\sigma = 0.23/\sigma = 0$ (cyan), $\sigma = 0.46/\sigma = 0$ (magenta). Look-up-tables of the images shown in (a-i) are independently scaled to the data shown.  These datasets can be accessed at\cite{Data}.}
      	\label{fig:beadExperiment}
	\end{figure}
        
        The effect of attenuation-compensation on the raw image intensity is clearly visible. As the strength of the compensation increases, beads deeper into the phantom (positive $x-$direction) become visible (Fig. \ref{fig:beadExperiment}(a,d,g)). Since the intensity is re-normalised in the deconvolved images, the intensity change is not visible, but an improved SBR (see Methods (Section \ref{sec:methods})) is observed with increasing $\sigma$ (Fig. \ref{fig:beadExperiment}(b,c,e,f,h,i)).
    
    	A spot-finding algorithm (see Methods (Section \ref{sec:methods})) was used to identify and measure individual beads in the data from Fig. \ref{fig:beadExperiment}(b,e,h). Figure \ref{fig:beadExperiment}(j,k) show the axial full-width at half-maximum (FWHM) and the local SBR for each compensation level, and Fig. \ref{fig:beadExperiment}(l,m) show ratios of the data relative to the case of $\sigma=0$. The lateral resolution for all compensation levels was $(1.2\pm0.2)\mum$ with no trend across the FOV. This is larger than expected for diffraction-limited performance, but consistent with a small amount of spherical aberration which was expected due to the walls of the capillary tube housing the sample \cite{Vettenburg2014}.
    
    \subsection{Attenuation-compensated Airy light-sheet microscopy in thick biological specimens}
	\label{subsec:BioResults}
    
    	Next, we tested the performance of attenuation-compensation in a range of thick biological specimens where attenuation is the result of a combination of absorption and scattering. In all specimens shown, the attenuation greatly exceeds the maximum that can be fully compensated across the FOV given the system parameters of our microscope (see Methods (Section \ref{sec:methods})).
    
    First, we imaged the cellular arrangement within the operculum of \textit{S. lamarcki}. The polychaete \textit{S. lamarcki} (formerly \textit{Pomatoceros lamarckii}) is a serpulid tubeworm that is found widely in temperate marine habitats, often in large numbers in rocky inter-tidal zones or biofouling man-made marine infrastructure. In recent years the animal has been used to better understand aspects of animal genome evolution and evolutionary developmental biology, including appendage regeneration \cite{Szabo2014}. The operculum is an appendage on the head of \textit{S. lamarcki}, used as a protective plug for the habitation tube when the animal is threatened, and is particularly accessible and amenable as a study system for regeneration. Improved methods of imaging this appendage are particularly valuable in this context.
    
    \begin{figure}[tb]
    	\centering
      	\includegraphics[width=\linewidth]{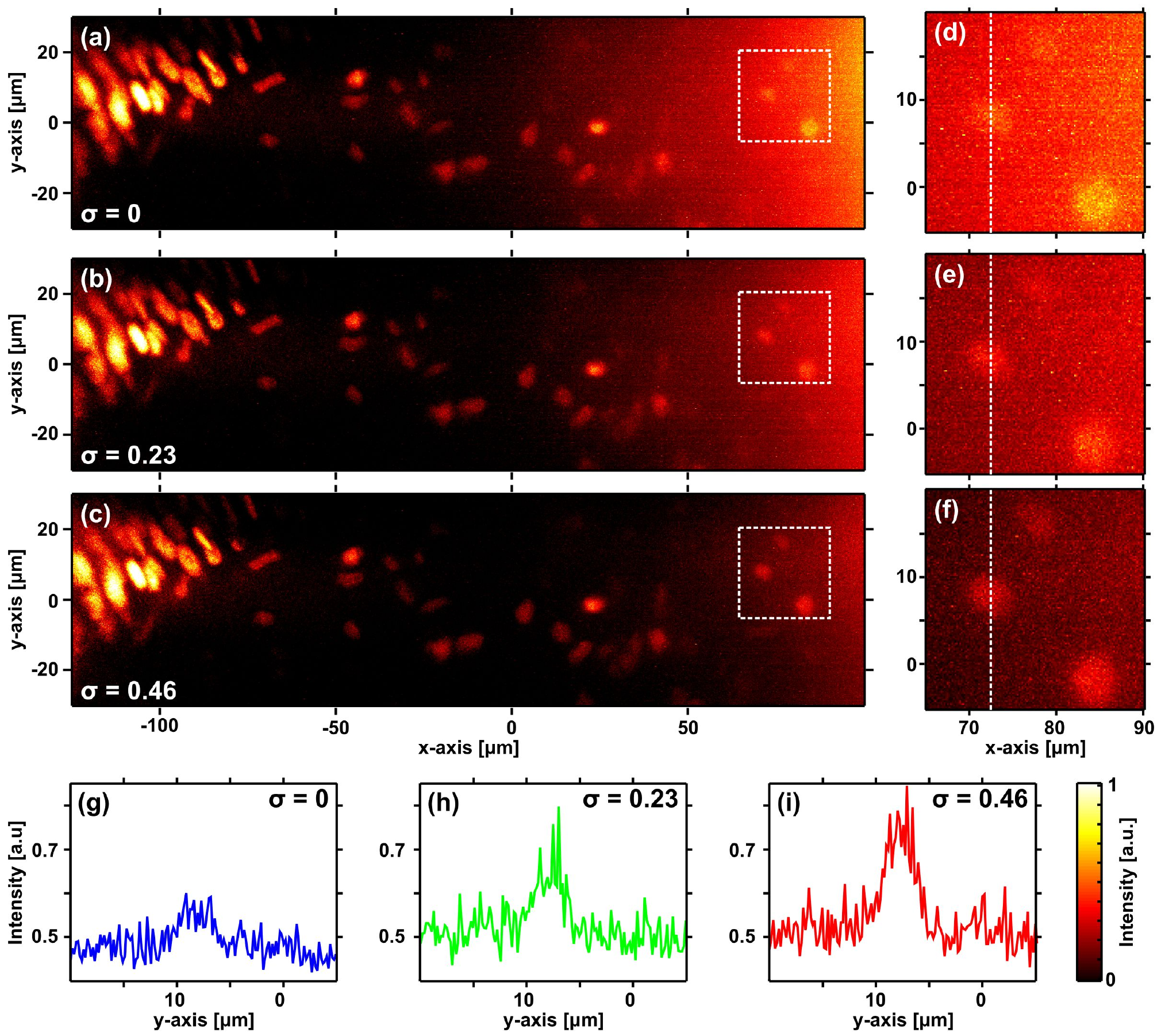}
      	\caption{Maximum intensity projections of deconvolved Airy LSM images of nuclei stained with propidium iodide in the operculum of \textit{S. lamarcki} (attenuation estimated at $85$cm$^{-1}$) with (a) no attenuation-compensation, (b) $\sigma = 0.23$, (c) $\sigma = 0.46$. (d-f) show expanded views of the region indicated by the dashed box in (a-c). (g-i) show intensity profiles along the dashed line shown in (d-f). These datasets can be accessed at\cite{Data}.}
      	\label{fig:SpirobranchusLamerki_Scan06_deepNuclei}
	\end{figure}
    
    Figure \ref{fig:SpirobranchusLamerki_Scan06_deepNuclei} compares images of the operculum at different levels of attenuation-compensation. In Fig. \ref{fig:SpirobranchusLamerki_Scan06_deepNuclei}(a), no compensation is applied ($C_{attn}' = 85$cm$^{-1}$). Compensation with $\sigma = 0.23$ ($C_{attn}' = 57$cm$^{-1}$) and $\sigma = 0.46$ ($C_{attn}' = 30$cm$^{-1}$) are shown in Fig. \ref{fig:SpirobranchusLamerki_Scan06_deepNuclei}(b,c). While full compensation cannot be achieved in this sample, increasing $\sigma$ steadily improves the SBR at depth in the tissue. Figure \ref{fig:SpirobranchusLamerki_Scan06_deepNuclei}(d-f) show expanded views of a region approximately $200\mum$ deep into the tissue, indicated by the dashed box in Fig. \ref{fig:SpirobranchusLamerki_Scan06_deepNuclei}(a-c). Without compensation, these deep nuclei are only marginally above the noise floor, but for $\sigma=0.46$ the SBR is increased between $20-45\%$ and the CNR is increased between $20-140\%$ (see Methods (Section \ref{sec:methods}) and Supplementary Fig. S20), raising features sufficiently above the noise floor for easy identification. Figure \ref{fig:SpirobranchusLamerki_Scan06_deepNuclei}(g-i) show the intensity profile along the dashed line in Fig. \ref{fig:SpirobranchusLamerki_Scan06_deepNuclei}(d-f), again highlighting the marked improvement in SBR and CNR with attenuation-compensation.
    
        We also compared the effect of attenuation-compensation to the `na\"{i}ve' case in which more intensity is delivered deeper into the tissue simply by increasing the overall illumination power at the back aperture of the objective to match that of the attenuation-compensated beam. Since high illumination intensity is linked to increased photobleaching and phototoxicity, lower illumination intensities are  preferable. In Fig. \ref{fig:SpirobranchusLamerki_Scan13}(a), a different region of the \textit{S. lamarcki} operculum is shown without attenuation-compensation ($C_{attn}=75$cm$^{-1}$). In Fig. \ref{fig:SpirobranchusLamerki_Scan13}(b), no attenuation-compensation is used, but the illumination power has been increased to match that of an attenuation-compensated light-sheet ($\sigma = 0.46$; Fig. \ref{fig:SpirobranchusLamerki_Scan13}(c); see Supplementary Table S3). Expanded views of the regions indicated by dashed boxes [i]-[iv] in Fig. \ref{fig:SpirobranchusLamerki_Scan13}(a-c) are shown in Fig. \ref{fig:SpirobranchusLamerki_Scan13}(d-o). Intensity profiles along the dashed line in Fig. \ref{fig:SpirobranchusLamerki_Scan13}(f,j,n) are shown in Fig. \ref{fig:SpirobranchusLamerki_Scan13}(p-r) respectively. In regions [ii]-[iv], the use on increased power without attenuation-compensation increased the SBR by $0-20\%$ and the CNR by $0-150\%$ whereas the use of attenuation-compensation increased the SBR by $20-40\%$ and the CNR by $50-650\%$ (see Supplementary Fig. S21).
        
        While increasing the power of the non-compensated beam (Fig. \ref{fig:SpirobranchusLamerki_Scan13}(b,h-k,q)) does improve the SBR and CNR at depth, it does so at the cost of increased irradiation of superficial tissues (see Supplementary Fig. S22), whereas the use of attenuation-compensation (Fig. \ref{fig:SpirobranchusLamerki_Scan13}(c,l-o,r)) increases the SBR and CNR at depth more, greatly aiding feature recognition, and does so with relatively uniform irradiation across the FOV (see Supplementary Fig. S22).

	\begin{figure}[tbp]
    	\centering
      	\includegraphics[width=\linewidth]{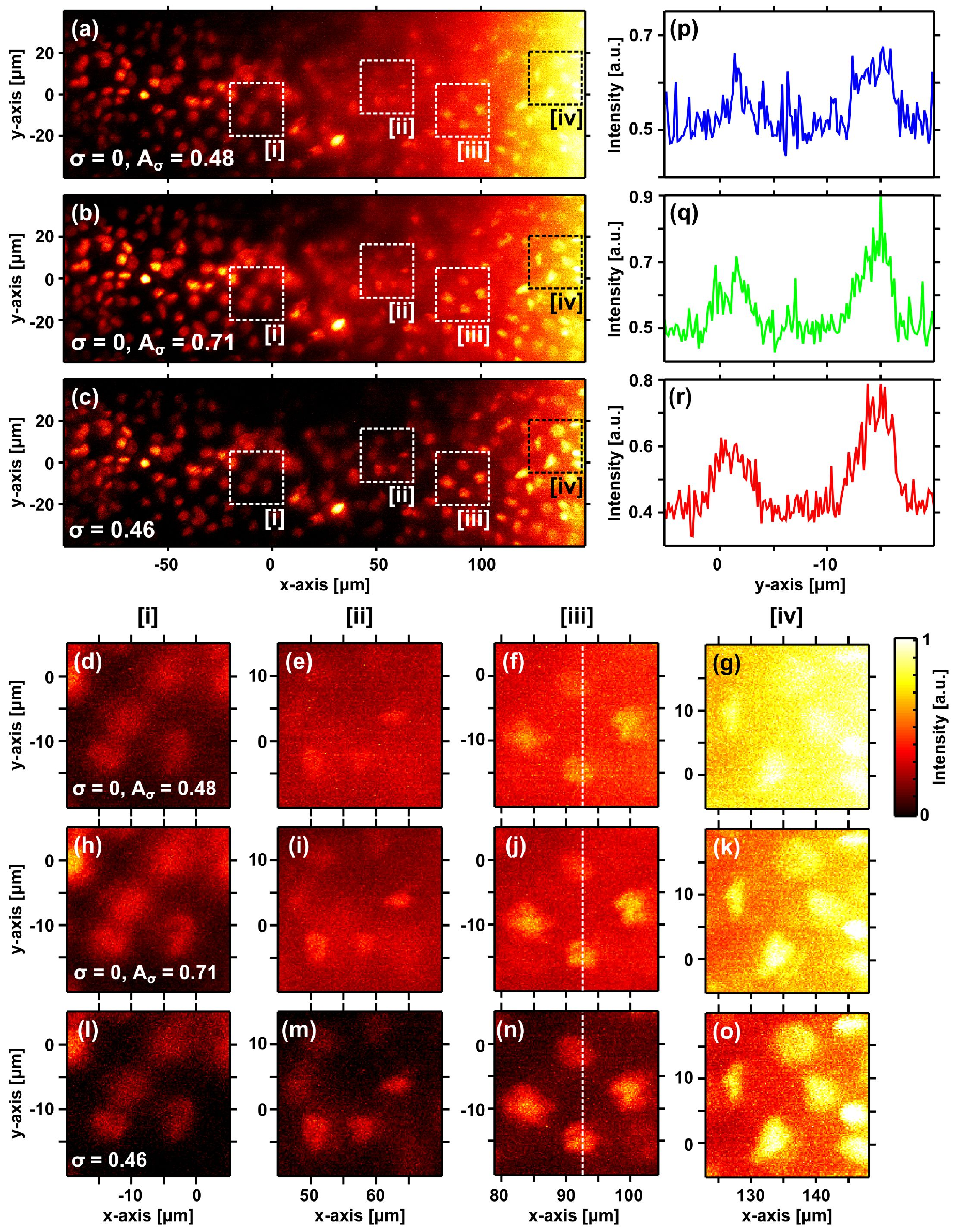}
      	\caption{Maximum intensity projections of deconvolved Airy LSM images of nuclei stained with propidium iodide in the operculum of \textit{S. lamarcki} (attenuation estimated at $75$cm$^{-1}$) with (a) no attenuation-compensation, (b) no attenuation-compensation but the same total power as a compensated light-sheet ($\sigma = 0.46$), (c) $\sigma = 0.46$. (d-g) show zoomed-in views of the region indicated by dashed boxes [i]-[iv] in (a), (h-o) show the same regions from (b,c). Intensity profiles along the dashed line in (f,j,n) are shown in (p-r) respectively. These datasets can be accessed at\cite{Data}.}
      	\label{fig:SpirobranchusLamerki_Scan13}
	\end{figure}

	Finally, we imaged fluorescently labelled kisspeptin neurons in the hypothalamic arcuate nucleus of mouse brain sections. Kisspeptins are a family of brain hormones that are necessary for pubertal development and the maintenance of fertility in humans and mice \cite{Topaloglu2012,Seminara2003}. Sex steroids dynamically control reproductive physiology through direct actions on kisspeptin neurons in the hypothalamus. One population of kisspeptin neurons located in the arcuate nucleus displays acute structural plasticity to circulating sex steroids; low sex steroid levels lead to cell hypertrophy whereas chronically high steroid levels leads to reduced cell size and decreased dendritic spine density \cite{Cholanian2015}. Studies have recently linked dysregulated kisspeptin neuronal signalling to increased accumulation of visceral fat, glucose intolerance and menopausal hot flushes, indicating that appropriate kisspeptin brain functions are important for preventing diseases not only associated with the reproductive system, but also diabetes, energy imbalance and thermoregulatory disorders. Improved methods for imaging the interaction of kisspeptin neurons throughout the brain will be beneficial for neuroendocrine research in particular and for neuroscience in general.

	We switched from a 2+1D Airy DSLM system to a 1+1D Airy SPIM system (see Methods (Section \ref{sec:methods})) in order to capture the long-range processes of kisspeptin neurons. Both systems are identical in operation and performance, except for the $y-$axis extent of the light-sheet which is at least 4 times larger in our SPIM system than in our DSLM system.

    \begin{figure}[tbp]
    	\centering
      	\includegraphics[width=\linewidth]{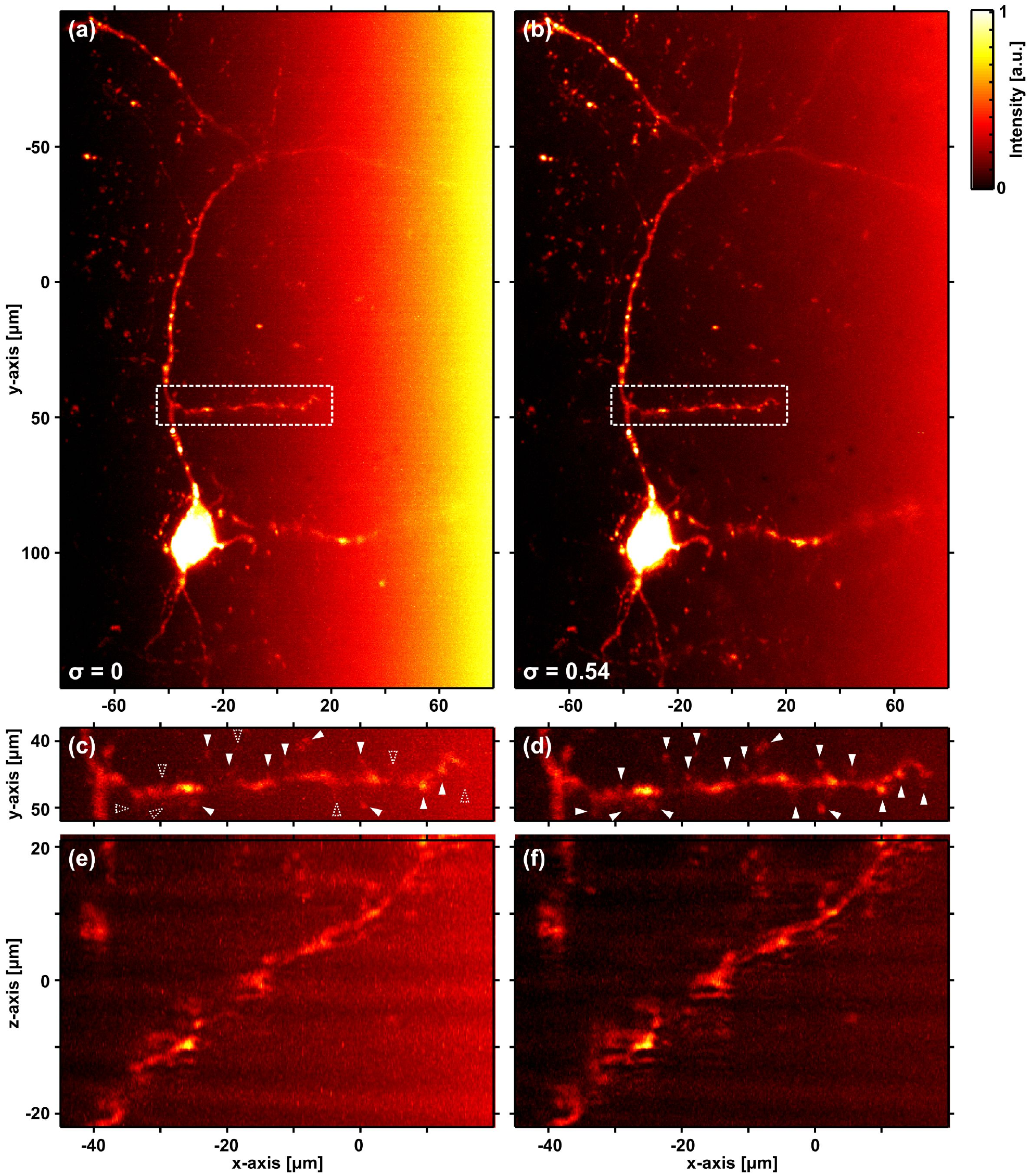}
      	\caption{Maximum intensity projections of deconvolved Airy LSM images of kisspeptin neurons expressing mCherry in the hypothalamic arcuate nucleus of a mouse brain (attenuation estimated at $100$cm$^{-1}$) with (a) no attenuation-compensation and (b) $\sigma = 0.54$. (c,d) show expanded views of the region indicated by the dashed box in (a,b). (e,f) show orthogonal projections of the regions shown in (c,d). Filled arrowheads indicate the positions of dendritic spines. Dashed arrowheads in (c) indicate the position of dendritic spines not identified without attenuation-compensation. These datasets can be accessed at\cite{Data}.}
      	\label{fig:MouseSPIM}
	\end{figure}
    
    Figure \ref{fig:MouseSPIM} shows an isolated soma and its processes. The attenuation of this tissue section has been estimated at $C_{attn} = 100$cm$^{-1}$. In Fig. \ref{fig:MouseSPIM}(a) no compensation has been applied. Attenuation-compensation ($\sigma = 0.54$, $C_{attn}' = 35$cm$^{-1}$) is applied in Fig. \ref{fig:MouseSPIM}(b). Figure \ref{fig:MouseSPIM}(c,d) show expanded views of the region indicated by the dashed box in Fig. \ref{fig:MouseSPIM}(a,b) and Fig. \ref{fig:MouseSPIM}(e,f) show orthogonal projections of this region.
    
    The increased contrast in the attenuation-compensated image allows a more accurate count of dendritic spines to be made. Filled arrowheads in Fig. \ref{fig:MouseSPIM}(c,d) indicate the positions of dendritic spines. Dashed arrowheads in Fig. \ref{fig:MouseSPIM}(c) indicate the position of dendritic spines not identified without attenuation-compensation. We manually identified 17 dendritic spines in Fig. \ref{fig:MouseSPIM}(d) but only 10 without attenuation-compensation (Fig. \ref{fig:MouseSPIM}(c)).

    \section{Discussion}
	\label{sec:discussion}
    
    Attenuation of the excitation light field is among the greatest limiting factors in the imaging of tissues and whole organisms in neuroscience and developmental biology. Recently, light-sheets generated from propagation-invariant Bessel and Airy beams that address the issue of scattering through their self-healing properties have come into use. Here, we have taken further advantage of the structure of the Airy light field to redistribute power according to an exponential increase with depth that compensates for absorption within the sample.
    
	Attenuation-compensation improved feature definition at depths of $50-200\mum$ in \textit{S. lamarcki} opercula and sections of mouse brain tissue, which we quantified using the local SBR and CNR. The maximum compensation applied in \textit{S. lamarcki} led to an SBR improvement of $20-45\%$ and CNR improvement of $50-650\%$. For images of large biological structures, such as nuclei in \textit{S. lamarcki}, CNR much more accurately quantifies the visual improvement seen in the images than SBR. In mouse brain tissue, attenuation-compensation enabled the clear identification of 7 additional dendritic spines (an extra $70\%$) that could not be distinguished from background without our approach (Fig. \ref{fig:MouseSPIM}(c,d)). Both the specimens used in this study had measured attenuation coefficients greater than the maximum we could compensate for, $C_{attn} = 65$cm$^{-1}$. However, in both cases, even partial attenuation-compensation resulted in dramatic improvements in image quality.

	We compared attenuation-compensation with the na\"{i}ve case of simply increasing the power of an Airy light-sheet without attenuation-compensation (Fig. \ref{fig:SpirobranchusLamerki_Scan13}). These solutions both achieve the delivery of more intensity to greater depths in the specimen, but when we consider the intensity distribution across the specimen (see Supplementary Fig. S22) we see that the intensity variation is minimised when using attenuation-compensation and restored close to the ideal (non-attenuated) case. While LSM in general minimises photobleaching and phototoxicity compared to other imaging modalities \cite{Keller2008,Jemielita2013,Laissue2017}, the importance of maintaining low peak intensity throughout the FOV has been highlighted and relatively small differences (a factor of 2-4) in intensity have shown a marked difference in photobleaching rate \cite{Vettenburg2014,Gao2012,Chen2014}. Attenuation-compensation in LSM may be an innovative route to low phototoxicity in deep-tissue imaging. Compared to multi-view methods, which achieve improved imaging of deep tissues through fusion of multiple images of the specimen taken from different viewpoints \cite{Swoger2007,Krzic2012,Medeiros2015,Wu2016}, attenuation-compensated Airy LSM is a single-view imaging method and is intrinsically $n$ times faster and has $n$ times lower specimen irradiation than a multi-view method requiring $n$ different views for reconstruction.
     
    Whilst our system utilises a dynamically reconfigurable spatial light modulator (SLM), allowing the strength of attenuation-compensation to be varied and tuned, all biological specimens exhibited attenuation in excess of the maximum compensation that can be applied in this way. Practically, in most studies, we may use a fixed optical element with no need to adjust $\sigma$. Use of tailored linear neutral density filters may be suitable for low-cost implementation of attenuation-compensation in an Airy LSM system. Due to the complex 2D amplitude and phase profile required to produce attenuation-compensated Bessel beams (see Supplementary Note S2), low-cost implementations for Bessel LSM techniques may not be as readily available.
    
    The proposed approach for attenuation-compensation of Airy LSM in the single-photon excitation regime counteracts the attenuation of the illumination. Attenuation-compensation may be even more crucial in the multi-photon excitation regime, as absorption losses dominate at the longer illumination wavelengths required. This is the subject of a future study. Due to the non-linear intensity dependence of two-photon excitation fluorescence signals, we anticipate the usable FOV of such techniques to reduce to less than half in the presence of strong attenuation (see Supplementary Note S9). Overall, this is a powerful, easy-to-implement approach that avoids the need for any determination of the transmission matrix of the tissue, aberration correction, is not restricted to any `point', and readily elevates the SBR, CNR, and information content from wide-field LSM using propagation invariant beams. We anticipate this will find significant uptake across a wide range of users in LSM and finally note that this approach may also be used for studies of propagation-invariant beams in optical manipulation, optical coherence tomography and other forms of imaging.

    \section{Methods}
	\label{sec:methods}
    
    \subsection{Attenuation-compensated Airy light-sheet microscope}
	\label{subsec:Setup}
    
    The Airy light-sheet microscope is similar to that described in \cite{Vettenburg2014} and can operate in either SPIM or DLSM modalities. In brief, we generated the attenuation-compensated Airy light-sheet via a spatial light modulator (SLM; Hamamatsu LCOS X10468-04). The excitation laser (Laser Quantum Finesse, 5W, 532nm) was expanded to overfill the active area of the SLM, programmed to display the appropriate phase mask (eq. \eqref{eq:MainTextAiryPupil1D}; see Supplementary Note. S1), modulated in the first-order of a blazed diffraction grating. The SLM is imaged onto an acousto-optic deflector (AOD; Neos AOBD 45035-3), and then onto the back aperture of the illumination objective lens (Nikon CFI Apo 40x/0.80 DIC, w.d. 3.5mm, water immersion). Pinholes are placed in each relay telescope at the Fourier plane to select only the first-order beams from the SLM and AOD. The SLM was used to holographically control the  numerical aperture (NA) of the illumination. For all experiments, $\NA=0.42$ was set to correspond to $u=1$ (see Supplementary Note S1) and $\alpha=7$ was used which yielded a maximum FOV of $328\mum$ \cite{Vettenburg2014}.
    
    The identical illumination and detection objective lenses are mounted in a "dual-inverted" geometry. Collected fluorescence was imaged onto an sCMOS camera (Hamamatsu Orca flash 4.0 v2) via a tube lens (Thorlabs, TTL200). A pinhole at the back aperture of the detection objective lens restricted the NA to 0.4.
    
    Data acquisition software was written in-house in LabVIEW, and data processing and deconvolution software, in MATLAB.
    
    \noindent \textbf{2+1D Airy DSLM.} For DSLM operation, the SLM generates the phase profile for a 2+1D Airy beam, the beam is spherically focussed into the sample with pupil function described by eq. (S7), and the light-sheet is formed by rapid modulation of the first-order deflection angle from the AOD. Due to the limited modulation range, the $y-$axis extent of the light-sheet is limited to $\approx100\mum$.
    
    \noindent \textbf{1+1D Airy SPIM.} For SPIM operation, the SLM generates the phase profile for a 1+1D Airy beam and a cylindrical lens (Thorlabs, ACY254-050-A) is inserted before the illumination objective to form a line focus at the back aperture. The light-sheet is then directly formed by cylindrical focussing with pupil function described by eq. S6. The $y-$axis extent of the light-sheet exceeded the FOV of the camera ($356\mum$).
    
    \subsection{Fluorescent attenuating phantoms}
	\label{subsec:prep_beadSample}
    
    Strongly attenuating (absorbing) phantoms containing resolution markers were made by adding 600nm diameter red fluorescent beads (Duke Scientific, R600) into 6.92mM Neutral Red dye ($\epsilon = (15.9 \pm 0.2)$mM$^{-1}$cm$^{-1}$). The suspension was mixed with an equal volume of 1\% low melting point agarose and injected into a square-profile borosilicate capillary (Hawksley, Vitrotube 8250-100), which was sealed at the ends with putty (Hawksley, Crystaseal). The final concentration of Neutral Red in the phantom was 3.46mM ($C_{abs} = (55 \pm 1)$cm$^{-1}$).
    
    \subsection{Image analysis}
    \label{subsec:image_analysis}
    
    \noindent \textbf{Spot-finding algorithm.} Our spot-finding algorithm was implemented in MATLAB. We raster-scanned through a deconvolved data cube and identified any pixel whose intensity was above a pre-set threshold value. We checked that this pixel was the maximum within a 19 x 19 pixel ($3$ x $3 \mum^{2}$) square, re-centered on the true maximum if necessary, and fitted a three-dimensional Gaussian function. Spots with FWHM outwith the range of $0.5-3\mum$ were deemed to be noise or not a single isolated bead, and discarded from the analysis.
    
    To increase the speed of the algorithm, we performed the raster scan on an $x-y$ maximum intensity projection, only scanning along the third dimension when a local maximum was detected. We also logged the ($x$,$y$,$z$) pixel coordinates of the centre of each spot and deleted any duplicates from the final list.
    
    \noindent \textbf{Signal-to-background ratio.} The local signal-to-background ratio (SBR) is given by:
    
    \begin{equation}
    \label{eq:SBR}
    	\textrm{SBR} = \mu_{s} / \mu_{b}
    \end{equation}
    
    \noindent where $\mu_{s}$ is the mean pixel value within a given region-of-interest (ROI) identified as a structure of interest or `signal' and $\mu_{b}$ is the mean pixel value within an ROI of equal size identified as containing no structures of interest or `background'. Typically the ROIs chosen were $3$ x $3 \mum^{2}$ squares.
    
    \noindent \textbf{Contrast-to-noise ratio.} The local contrast-to-noise ratio (CNR) is given by:
    
    \begin{equation}
    \label{eq:CNR}
    	\textrm{CNR} = \frac{\mu_{s} - \mu_{b}}{\sqrt{\sigma_{s}^{2} + \sigma_{b}^{2}}}
    \end{equation}
    
    \noindent where $\mu_{s}$ and $\mu_{b}$ are the mean pixel values in `signal' and `background' ROIs as described above and $\sigma_{s}$ and $\sigma_{b}$ are the standard deviation of the pixel values in the `signal' and `background' ROIs respectively \cite{Adler2004}. Typically the ROIs chosen were $3$ x $3 \mum^{2}$ squares.
    
    \subsection{\textit{S. lamarcki} opercula}
    \label{subsec:prep_spirobranchus_lamarcki}
    
    \textbf{Animals.} Adults of polychaete \textit{Spirobranchus lamarcki} (formerly \textit{Pomatoceros lamarckii}) were collected at East Sands, St Andrews, and maintained in the circulating seawater aquarium system of the Scottish Oceans Institute, Gatty Marine Laboratory, at close to ambient seawater temperature.
    
    \noindent \textbf{Tissue collection.} Adults were removed from their calcareous habitation tube by crushing the posterior end of the tube until the opening was wide enough for the animal to pass through, and then using a blunt probe to push the animal out of the remaining tube from the anterior end. Animals were then rinsed in fresh filtered seawater and opercula removed with a scalpel, cutting at the Easy Break Point \cite{Szabo2014}. Isolated opercula were rinsed in filtered seawater and then as much of the water was removed as possible before fixing in $4\%$ paraformaldehyde (PFA) in 1x phosphate-buffered saline (1x PBS). The opercula were gently rotated in the fixative for 5 minutes before replacing with fresh fixative solution and fixing overnight at $4^{\circ}$C. The fixed opercula were then given two washes in $70\%$ ethanol, with gentle rotation at room temperature (RT), for 5 minutes each wash. This was repeated with absolute ethanol for two further 5 minute washes. The opercula were then kept in $70\%$ ethanol until used for propidium iodide (PI) staining.
    
    \noindent \textbf{Fluorescent labelling.}  For labelling with PI, half of the $70\%$ ethanol was replaced with PBT (1x PBS + $0.1\%$ Triton X100) and gently rotated for 5 minutes at RT. This step was repeated, and then all of the solution was replaced with PBT and rotated gently for 5 minutes at RT. The opercula were then washed in fresh PBT with gentle rotation for at least 1 hour. Two 2x saline sodium citrate (2x SSC) washes of 5 minutes at RT with gentle rotation followed, prior to treatment with $100\upmu$g/mL RNAseA in 2x SSC for 20 minutes at $37^{\circ}$C. The RNAseA was removed by three 5-minute washes with 2x SSC at RT with gentle rotation, and the opercula stained with PI, using a 1:1000 dilution of a 1mg/mL stock diluted in 2x SSC. The staining was carried out for 30 minutes in the dark at RT, and excess PI then removed by three 5-minute washes in 2x SSC at RT with gentle rotation in the dark. Samples were kept in 2xSSC in the dark at $4^{\circ}$C until imaged.
    
    \noindent \textbf{Imaging.} For imaging, samples were placed on a microscope slide, embedded in $1\%$ low melting point agarose, and immersed in 1x PBS on the microscope sample stage.
    
    \subsection{Mouse brain sections}
    \label{subsec:prep_mouse_brain_section}

	\textbf{Animals.} All rodent experiments were reviewed and approved by the University of St Andrews Animal Ethics and Welfare Committee under Dr Tello's Home Office Project License 70/7924. Adult female heterozygous \textit{Kiss1}$^{\text{tm1.1(cre/EGFP)Stei}}$/J mice (10-12 weeks, 20-25g) were used for this study. Mice were housed in a temperature- and humidity-controlled environment under regular light-dark cycles (12h light, 12h dark) with food and water available \textit{ad libitum}.
    
    \noindent \textbf{\textit{In vivo} AAV infusion.} Mice were anaesthetised with isofluorane and placed in a sterotaxic apparatus. To selectively express mCherry (a monomeric red fluorescent protein) in arcuate \textit{Kiss1$^{+}$} neurons, we used Cre recombinase (Cre)-dependent adenoassociated virus vector \cite{McClure2011} (AAV; AAV1/2-Ef1a-DIO-mCherry-wPRE as described previously \cite{Nylk2016}). Viral particles were injected bilaterally into the hypothalamic arcuate nucleus (coordinates: AP -1.7, ML $\pm$0.3, DV -5.9) using a pulled glass pipette at a volume of 400nL/side, at a rate of 100nL/min using pressure injection. After surgery, mice were returned to their cages for 3 weeks to allow for AAV-mediated mCherry expression.
    
    \noindent \textbf{Perfusion and tissue sectioning.} After 3 weeks, mice were administered with an overdose of sodium pentobarbital (100mg/kg) and transcardially perfused with $0.1$M PBS (pH $7.4$) followed by $4\%$ PFA in PBS (pH $7.4$). Brains were removed from the skull and post-fixed overnight in $4\%$ PFA in PBS and subsequently cryopreserved in $30\%$ sucrose in $0.1$M PBS. The brains were sectioned using a Compresstome vibratome (Precisionary Instruments VF-300) at a thickness of $300\mum$.
    
    \noindent \textbf{Imaging.} For imaging, tissue sections were placed on a microscope slide, embedded in $1\%$ low melting point agarose, and immersed in 1x PBS on the microscope sample stage.

    \section*{Acknowledgements}
    
    We thank the UK Engineering and Physical Sciences Research Council for funding through grants (EP/P030017/1	and EP/J01171X/1). J.A.T. acknowledges funding from the British Society for Neuroendocrinology Project Support Grant as well as the RS MacDonald Trust. D.E.K.F. acknowledges funding from the Leverhulme Trust. We thank Mingzhou Chen for useful discussions on image analysis.

    \section*{Author contributions}
    
    J.N. and K.D. directed the study. J.N., M.P., and M.M. developed theoretical aspects of the study. J.N. performed the simulations. J.N., and K.M. performed all experiments, data processing, and analysis. D.E.K.F. provided and prepared \textit{S. lamarcki} samples, and assisted with imaging and interpretation. S.A., and J.A.T. provided and prepared mouse brain samples, and assisted with imaging and interpretation. F.J.G.-M. assisted with the biological experiments. J.N., K.M. and K.D. wrote the manuscript with contributions from the other authors. K.D. initiated and supervised the work.


    \endgroup

	\end{document}